\documentclass{article}
\usepackage{hiph-art}
\usepackage{epsfig}
\volnumber{22} \issuenumber{1} \edyear{2005}                             
\frompage{000} \topage{000}                                              
\recrevdate{19 October 2005}                                              
\def\rightmark{Freeze our in narrow and wide layers}
\def\leftmark{E. Moln\'ar et al.}

\def\be{\begin{equation}}
\def\ee{\end{equation}}
\def\bea{\begin{eqnarray}}
\def\eea{\end{eqnarray}}

\title{Covariant kinetic freeze out description through \\ a finite space-time layer}

\authors{
{E. Moln\'ar$^{1}$, L.P. Csernai$^{1}$ and  V.K. Magas$^{2}$ %
\index{E. Moln\'ar et al.} 
}\\[2.812mm]
{\normalsize
\hspace*{-8pt}$^1$ Section for Theoretical and Computational Physics,\\
     University of Bergen, Allegaten 55, 5007 Bergen, Norway\\
\hspace*{-8pt}$^2$~Departament d'Estructura i Constituents de la Mat\'eria, \\
Universitat de Barcelona, Diagonal 647, 08028 Barcelona, Spain\\[0.2ex]
}}

\abstract{
The problem of Freeze Out (FO) in high energy heavy ion reactions is addressed.
We develop and analyze a covariant FO description valid for a finite space-time layer.
}


\makeindex

\begin{document}
\maketitle
\section{Introduction}

Freeze out (FO) is a term referring to the stage of expanding or exploding matter when its constituents
(particles) lose contact, collisions cease, and local dynamical equilibrium is not maintained.
In local equilibrium the evolution of the system can be described by hydrodynamics, while
as time passes the system becomes more dilute, the number of non-interacting particles
increases until the whole system breaks up and the particle momentum is frozen out.

\section{Freeze out and the Boltzmann Transport Equation}

The conservation of the total number of particles with collisions among them requires
the relativistic Boltzmann Transport Equation (BTE),
\begin{equation}\label{BTE}
p^\mu \partial_\mu f =
 \frac{1}{2}\int {_{12}\mathcal{D}}_4    f_1 f_2 W_{12}^{p4}
-\frac{1}{2}\int {   _2\mathcal{D}}_{34} f   f_2 W_{p2}^{34} \, ,
\end{equation}
commonly written in form of gain and loss terms.
We have used the following shorthand notation for the invariant scalar product
${_{ab}\mathcal{D}}_c  \equiv \frac{d^3 p_a}{p_a^0}\frac{d^3 p_b}{p_b^0} \frac{d^3 p_c}{p_c^0}$.
Furthermore, the dynamics is governed by the invariant transition rates, $W_{ab}^{cd}$,
which stands for the elementary reaction, $a + b \rightarrow c + d$, satisfying the energy
momentum conservation, while $f_j \equiv f(x,p_j)$.
\\ \indent
To describe a gradual freeze out process, we split the $f$ distribution function into two
parts: $f=f^i + f^f$.
The result of collisions is the drain of particles from the interacting component, $f^i$, which
gradually builds up the free component, $f^f$, expressed by the freeze out probability,
$\mathcal{P}_f$.
The FO probability populates the free component, while the rest, $(1 - \mathcal{P}_f)$,
feeds the interacting component.
We can separate the two components into two equations:
\begin{eqnarray}\label{eq_int}
p^\mu \partial_\mu f^f \! \! \! &=& \!\! \!
+ \, \frac{1}{2}\int {_{12}\mathcal{D}}_4 f^i_1 f^i_2 \ {\mathcal{P}}_f W_{12}^{p4} \, ,\\ \nonumber
p^\mu \partial_\mu f^i  \! \!\! &=&\! \! \!
- \, \frac{1}{2}\int {_{12}\mathcal{D}}_4    f^i_1 f^i_2 {\mathcal{P}}_f W_{12}^{p4}
+ \frac{1}{2}\int {_{12}\mathcal{D}}_4    f^i_1 f^i_2 W_{12}^{p4}
- \frac{1}{2}\int {   _2\mathcal{D}}_{34} f^i   f^i_2 W_{p2}^{34} \, ,
\end{eqnarray}
so that the sum of these two equations returns eq. (\ref{BTE}).
The free component does not contain a loss term because the free component
cannot loose particles due to collisions.
The latter two terms in the interacting component do not include $\mathcal{P}_f$
as they influence only the interacting term by redistributing particles in the momentum space and
driving the interacting component towards rethermalization \cite{ModifiedBTE}.

\section{Approximate kinetic freeze out equations}

\begin{figure}[!htb]
\centering
\includegraphics[width=5.6cm, height = 4.4cm]{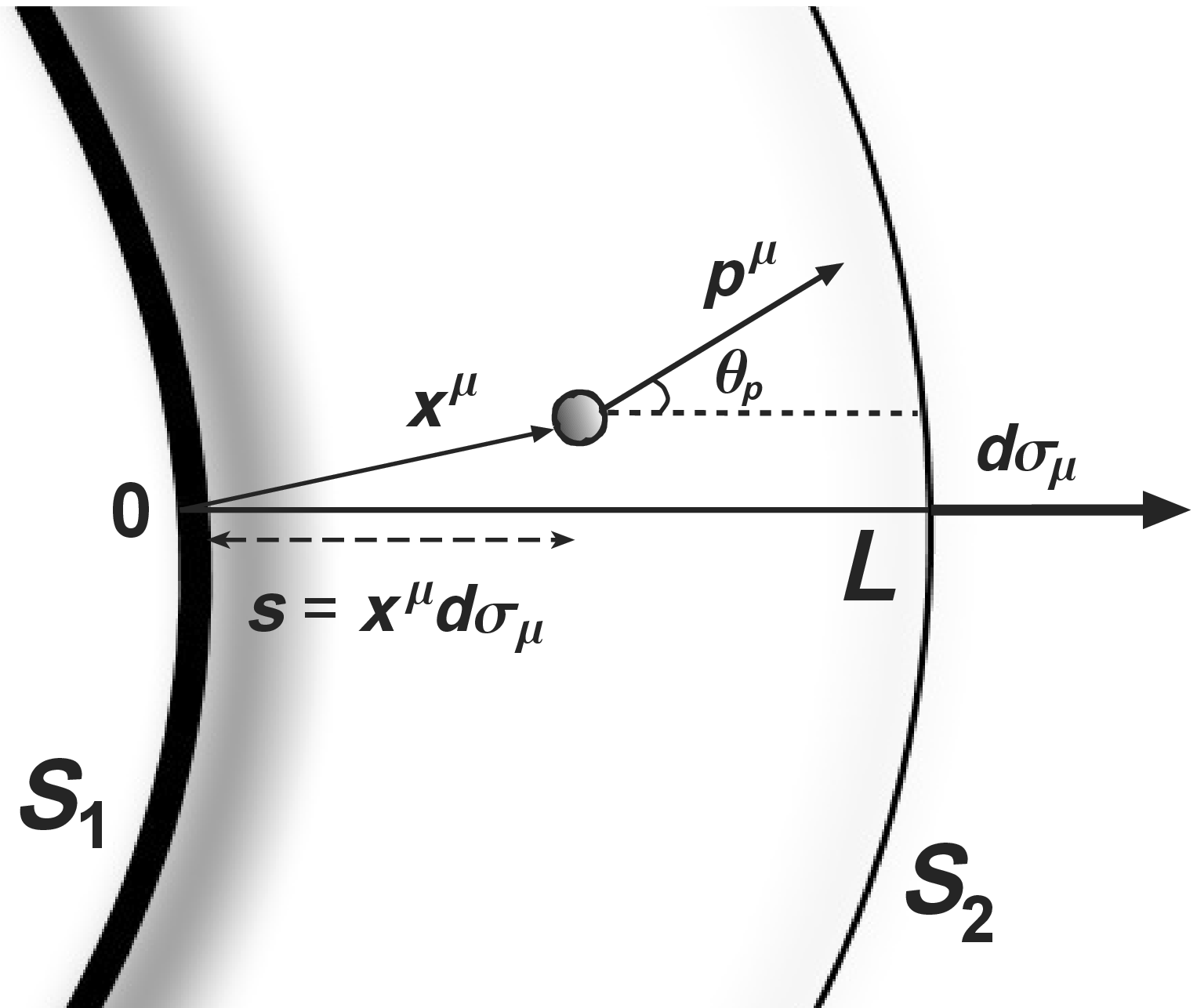}
\includegraphics[width=5.6cm, height = 4.4cm]{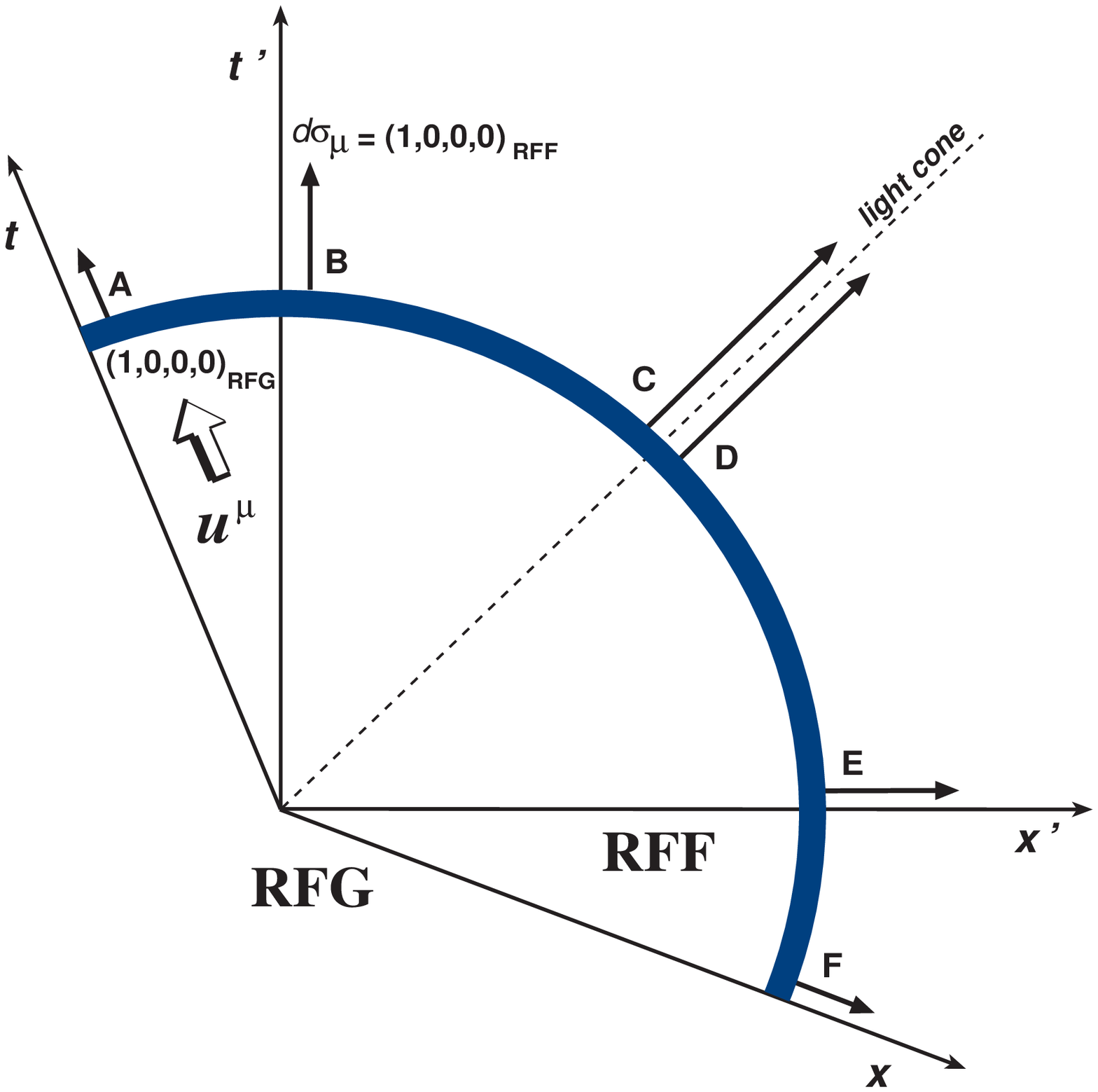}
\caption{Left: The picture of a gradual FO process within the finite FO layer, in coordinate space.
The inside boundary of the FO layer, $S_1$, indicates the points
where the FO starts. This is the origin of the coordinate vector, $x^{\mu}$.
Within the finite thickness of the FO layer, $L$, the density of the
interacting particles gradually decreases and disappears at the outside boundary
$S_2$ of the FO layer.
Right: A simple hypersurface in the Rest Frame of the Gas (RFG:\,[t,x]) ,
where $u^{\mu} = (1,0,0,0)$, including time-like and space-like parts.
On these two parts of the hypersurface in the Rest Frame of the Front (RFF:\,[t',x']),
the normal to the hypersurface, $d\sigma_{\mu}$, points into the direction of the $t'$ ($x'$)-axis
respectively.} \label{updatedFO}
\end{figure}

\begin{figure}[!hb]
\centering
\includegraphics[width=10.2cm, height = 4.4cm]{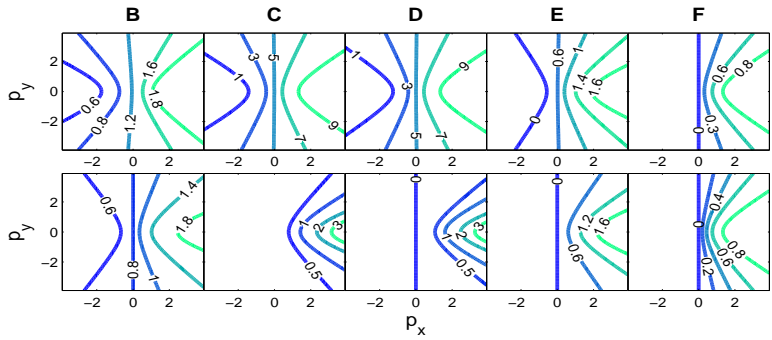}
\caption{ The contour plots of the momentum dependent part of the
escape rate, $W(p)$, presented at different points of the FO hypersurface.
The upper plots show the values of $W(p)$ in RFG, while the plots below are
in RFF. The momenta are in units of particle mass.}
\label{ps_plot}
\end{figure}

We do not possess enough information to calculate the gain term containing the FO probability
from eq. (\ref{eq_int}), thus we rather approximate this term based on fundamental physical
principles with the so called escape rate, $\mathcal{P}_{esc}$.
The escape rate includes $\mathcal{P}_f$, and separates the outgoing (gain) particles
into a fraction that is still colliding and a fraction that is not.
\\ \indent
The probability not-to-collide with anything on the way out, should depend on the number
of particles, which are in the way of a particle moving outwards in the direction $p_x/|\vec{p}|$
across the FO layer of thickness $L$.
Following a particle moving outwards form the beginning, i.e. ($x^{\mu}=0$),
to a position $x^\mu$, with momentum $p^{\mu}$, the actual distance up to the outer FO layer is
$(L-s)/ \cos \theta_{\vec{p}} = (L - x^\mu d\sigma_\mu)/\cos \theta_{\vec{p}}$, where
$s=x^{\mu}d\sigma_{\mu}$, see Fig. \ref{updatedFO}.
Assuming that the FO probability is inversely proportional to the remaining distance,
the escape rate is:
\be\label{escb}\nonumber
\mathcal{P}_{esc} \equiv
\frac{1}{\lambda} \left( \! L\times\frac{\cos \theta_{\vec{p}}}{L - x^\mu d\sigma_\mu} \right)
\!\Theta(p^\mu d\sigma_\mu)\, ,
\ee
where the cut-off factor, $\Theta(p^{\mu}d \sigma_{\mu})$, eliminates
the FO of particles with negative momentum and $\lambda$ is the initial characteristic length of the system.
If the particle momentum is not normal to the surface, the remaining spatial distance
further increases with increasing $\theta_{\vec{p}}$ and the probability to
leave the system decreases as $\cos\theta_{\vec{p}}$, where $\theta_{\vec{p}}$ is
the angle between the FO normal vector and $\vec{p}$, see Fig. \ref{updatedFO}.
This "naive" rate, $\cos\theta_{\vec{p}}$, is not covariant, and did not take into account that the
escape rate of particles is also proportional to the particle velocity, $|\vec{v}|$.
The covariant generalization can be given in the following way \cite{articles_1_2}:
\be\label{escape_rate}
\cos\theta_{\vec{p}} \times |\vec{v}| \equiv \frac{p^{\mu}}{|\vec{p}|}\times\frac{|\vec{p}|}{p^{0}}
\Rightarrow \bigg( \frac {p^\mu d\sigma_\mu}{p^\mu u_\mu} \bigg) \, ,
\ee
where $p^{\mu}=(p^0, \vec{p})$ is the particle 4-momentum , $u^{\mu}$ is the flow velocity.
Now, using the invariant form of the escape rate, the approximate equations governing
the FO development for both time-like and space-like FO situations are:
\bea \label{general}
d\sigma^{\mu} \partial_{\mu} f^{f} \! &=& \! + \frac{1}{\lambda}
\bigg(\!\frac{L}{L - x^{\mu} d\sigma_{\mu}} \!\bigg)
\! \bigg( \! \frac{p^{\mu} d\sigma_{\mu}}{p^{\mu} u_{\mu}} \! \bigg) \! \Theta(p^{\mu}d\sigma_{\mu}) f^{i} \, ,
\\ \nonumber
d\sigma^{\mu} \partial_{\mu} f^{i} \! &=& \! - \frac{1}{\lambda} \bigg(\!\frac{L}{L - x^{\mu} d\sigma_{\mu}}\! \bigg)
\! \bigg( \! \frac{p^{\mu} d\sigma_{\mu}}{p^{\mu} u_{\mu}} \! \bigg) \! \Theta(p^{\mu}d\sigma_{\mu}) f^{i}
+ \big[ f^{i}_{eq} - f^{i} \big]\frac{1}{\lambda_{th}} \, ,
\eea
where the equations describe a space-like, $d\sigma^{\mu}d\sigma_{\mu} = -1$, or a time-like,
$d\sigma^{\mu}d\sigma_{\mu} = 1$, FO process depending on the FO normal vector.
The equilibrium distribution function is denoted by $f^{i}_{eq}$, while $\lambda_{th}$ is
the rethermalization length or time.
\\ \indent
We can study this new angular factor, $W(p)=p^{\mu}d\sigma_{\mu}/p^{\mu}u_{\mu}\Theta(p^{\mu}d\sigma_{\mu})$,
by taking different typical points of the FO hypersurface.
At points A, B, C, the hypersurface is time-like, while at points
D, E, F, the hypersurface is space-like.
The resulting phase-space escape rates are shown in Fig. \ref{ps_plot}
for the cases B, C, D, E, F, in the Rest Frame of the Gas (RFG)
as well as in the Rest Frame of the Front (RFF).
For point A on the hypersurface, $W(p)=1$, in both reference frames.
\\ \indent
The effect of this relativistically invariant angular factor leads to a smoothly changing behavior of
$W(p)$, as the direction of the normal vector changes in RFG.
In RFF, $W(p)$ behaves discontinuously when we cross the light cone,
from point C to point D, (see the contour line belonging to $W(p)=0$).
This is a consequence the of the chosen reference frame only \cite{kemer,articles_1_2}.
\\
\\ \indent
The aim of freeze out calculations is to find the post FO momentum distribution and the
relevant quantities from the properties of the matter on the pre FO side.
The final post FO particle distributions are non-equilibrated and anisotropic
distributions.
These distributions in general cannot be Lorentz transformed to a frame where, the distribution
is isotropic \cite{articles_1_2,magas}.
The only exception is when the normal to the FO hypersurface is parallel to the local flow velocity.
The usual practice of assuming the J\"uttner or the cut-J\"uttner distribution as the post
FO distribution is generally not valid.
By introducing a finite thickness FO layer, we are strongly affecting the evolution
of both the interacting and frozen out components \cite{articles_1_2,magas}.


\end{document}